\RequirePackage[2020-02-02]{latexrelease}
\documentclass[twocolumn,preprintnumbers]{revtex4}%
\usepackage{amssymb}
\usepackage{amsmath}
\usepackage{graphicx}
\usepackage{dcolumn}
\usepackage{bm}
\usepackage{xcolor}
\usepackage{ifthen}
\usepackage[normalem]{ulem}
\usepackage{amsfonts}%
\RequirePackage[2020-02-02]{latexrelease}
\UseRawInputEncoding
\providecommand{\U}[1]{\protect\rule{.1in}{.1in}}
\providecommand{\U}[1]{\protect\rule{.1in}{.1in}}
\def\showal{1}
\newcommand{\al}[1]{\ifthenelse{\showal=1}{\textcolor{orange}{[[#1]]}}{}}

\newcommand{\eb}[1]{\ifthenelse{\showal=1}{\textcolor{cyan}{[[#1]]}}{}}
\begin{document}
\title{Partial disentanglement in a multipartite system}
\author{Eyal Buks}
\email{eyal@ee.technion.ac.il}
\affiliation{Andrew and Erna Viterbi Department of Electrical Engineering, Technion, Haifa
32000, Israel}
\date{\today }

\begin{abstract}
We explore a nonlinear extension to quantum theory giving rise to
deterministic partial disentanglement between pairs of subsystems. The
extension is based on a modified Schr\"{o}dinger equation having an added
nonlinear term. To avoid conflicts with the principles of causality and
separability, it is postulated that disentanglement is active only during the
time when particles interact. A butterfly-like effect is found near highly
entangled multipartite vector states.

\end{abstract}
\maketitle





\textbf{Introduction} - The problem of quantum measurement
\cite{Schrodinger_807} arguably originates from a self-inconsistency in
quantum theory \cite{Penrose_4864,Leggett_939,Leggett_022001}. In an attempt
to resolve this long-standing problem, several nonlinear extensions to quantum
theory \cite{Geller_2111_05977} have been proposed
\cite{Weinberg_336,Weinberg_61,Doebner_397,Doebner_3764,Gisin_5677,Kaplan_055002,Munoz_110503}%
, and processes giving rise to spontaneous collapse have been explored
\cite{Bassi_471,Pearle_857,Ghirardi_470,Bassi_257,Bennett_170502,Kowalski_1,Fernengel_385701,Kowalski_167955}%
. For some cases, however, nonlinear quantum dynamics may give rise to
conflicts with well-established physical principles, such as causality
\cite{Bassi_055027,Jordan_022101,Polchinski_397,Helou_012021,Rembielinski_012027,Rembielinski_420}
and separability \cite{Hejlesen_thesis,Jordan_022101,Jordan_012010}. In
addition, some predictions of standard quantum mechanics, which have been
experimentally confirmed to very high accuracy, are inconsistent with some of
the proposed extensions.

A nonlinear mechanism giving rise to suppression of entanglement (i.e.
disentanglement) has been recently proposed \cite{Buks_2303_00697}. This
mechanism of disentanglement, which makes the collapse postulate redundant, is
introduced by adding a nonlinear term to the Schr\"{o}dinger equation. The
proposed modified Schr\"{o}dinger equation can be constructed for any physical
system whose Hilbert space has finite dimensionality, and it does not violate
norm conservation of the time evolution. The nonlinear term added to the
Schr\"{o}dinger equation has no effect on product (i.e. disentangled) states.

The derivation of the added term for the case of a system composed of two
subsystems is given in Ref. \cite{Buks_2303_00697}. A system belonging to this
class is henceforth referred to as a bipartite. The main purpose of the
current paper is to propose a generalization, applicable for the case where
the system under study is divided into more than two subsystems (the
multipartite case). The derivation of the nonlinear term added to the
Schr\"{o}dinger equation for the multipartite case, which is discussed below,
is related to the quantification problem of subsystems' quantum entanglement
\cite{Schlienz_4396,Peres_1413,Hill_5022,Wootters_1717,Coffman_052306,Vedral_2275,Eltschka_424005,Dur_062314,Coiteux_200401,Takou_011004,Elben_200501}%
.

\textbf{Partial disentanglement} - Consider a modified Schr\"{o}dinger
equation for the ket vector $\left\vert \psi\right\rangle $ having the form%
\begin{equation}
\frac{\mathrm{d}}{\mathrm{d}t}\left\vert \psi\right\rangle =\left[
-i\hbar^{-1}\mathcal{H}-\gamma\left(  \mathcal{Q}-\left\langle \mathcal{Q}%
\right\rangle \right)  \right]  \left\vert \psi\right\rangle \;, \label{MSE}%
\end{equation}
where $\hbar$ is the Planck's constant, $\mathcal{H}^{{}}=\mathcal{H}^{\dag}$
is the Hamiltonian, the rate $\gamma$ is positive, the operator $\mathcal{Q}$
is allowed to depend on $\left\vert \psi\right\rangle $, and $\left\langle
\mathcal{Q}\right\rangle \equiv\left\langle \psi\right\vert \mathcal{Q}%
\left\vert \psi\right\rangle $. Note that the norm conservation condition
$0=\left(  \mathrm{d}/\mathrm{d}t\right)  \left\langle \psi\right.  \left\vert
\psi\right\rangle $ is satisfied by the modified Schr\"{o}dinger equation
(\ref{MSE}), provided that $\left\vert \psi\right\rangle $ is normalized, i.e.
$\left\langle \psi\right.  \left\vert \psi\right\rangle =1$. As is
shown in appendix \ref{App_tau}, the added nonlinear term proportional to
$\gamma$ in Eq. (\ref{MSE}) suppresses the expectation value $\left\langle
\mathcal{Q}\right\rangle $ (provided that $\gamma$ is positive). In
particular, this term can be constructed to suppress entanglement, i.e. to
give rise to disentanglement. The case of a bipartite system was
discussed in \cite{Buks_2303_00697}. Here we consider the more general case of
a multipartite system, and derive a modified Schr\"{o}dinger equation having
the form of Eq. (\ref{MSE}), which can give rise to partial disentanglement
between any pair of subsystems.

Consider a multipartite system composed of three subsystems labeled as '1',
'2' and '3', respectively. The Hilbert space of the system $H=H_{3}\otimes
H_{2}\otimes H_{1}$ is a tensor product of subsystem Hilbert spaces $H_{1}$,
$H_{2}$ and $H_{3}$. The dimensionality of the Hilbert space $H_{n}$ of
subsystem $n$, which is denoted by $d_{n}$, where $n\in\left\{  1,2,3\right\}
$, is assumed to be finite. The system is assumed to be in a pure state, which
is denoted by $\left\vert \psi\right\rangle $.

For any given observable $A_{1}^{{}}=A_{1}^{\dag}$ of subsystem 1, and a given
observable $A_{2}^{{}}=A_{2}^{\dag}$ of subsystem 2, the operator
$\mathcal{C}_{12}\left(  A_{1},A_{2}\right)  $ is defined by%
\begin{equation}
\mathcal{C}_{12}\left(  A_{1},A_{2}\right)  =A_{2}A_{1}\left\vert
\psi\right\rangle \left\langle \psi\right\vert -A_{1}\left\vert \psi
\right\rangle \left\langle \psi\right\vert A_{2}\;.
\end{equation}
Subsystems 1 and 2 are said to be disentangled if $\left\langle \mathcal{C}%
_{12}\left(  A_{1},A_{2}\right)  \right\rangle =0$ for any single particle
observables $A_{1}$ and $A_{2}$. A general single particle observable of
subsystem $n$, where $n\in\left\{  1,2\right\}  $, can be expanded using the
set of generalized Gell-Mann matrices $\left\{  \lambda_{1}^{\left(  n\right)
},\lambda_{2}^{\left(  n\right)  },\cdots,\lambda_{d_{n}^{2}-1}^{\left(
n\right)  }\right\}  $, which spans the SU($d_{n}$) Lie algebra. The
entanglement between subsystems 1 and 2 can be quantified by the nonnegative
variable $\tau_{12}$, which is given by $\tau_{12}=\left\langle \mathcal{Q}%
_{12}\right\rangle $, where the operator $\mathcal{Q}_{12}$ is given by
[compare to Eq. (31) of Ref. \cite{Schlienz_4396}]%
\begin{equation}
\mathcal{Q}_{12}=\eta_{12}\sum_{a_{1}=1}^{d_{1}^{2}-1}\sum_{a_{2}=1}%
^{d_{2}^{2}-1}\mathcal{C}_{12}\left(  \lambda_{a_{1}}^{\left(  1\right)
},\lambda_{a_{2}}^{\left(  2\right)  }\right)  \left\vert \psi\right\rangle
\left\langle \psi\right\vert \mathcal{C}_{12}\left(  \lambda_{a_{1}}^{\left(
1\right)  },\lambda_{a_{2}}^{\left(  2\right)  }\right)  \;, \label{Q12}%
\end{equation}
and where $\eta_{12}$ is a positive constant.

In a similar way, the entanglement between subsystems 2 and 3, which is
denoted by $\tau_{23}$, and the entanglement between subsystems 3 and 1, which
is denoted by $\tau_{31}$, can be defined. The completeness relation that is
satisfied by the generalized Gell-Mann matrices [see Eq. (8.166) of
\cite{Buks_QMLN}] can be used to show that $\tau_{23}$ , $\tau_{31}$\ and
$\tau_{12}$ are all invariant under any single subsystem unitary
transformation \cite{Schlienz_4396}. Deterministic disentanglement between
subsystems $n^{\prime}$ and $n^{\prime\prime}$ can be generated by the
modified Schr\"{o}dinger equation (\ref{MSE}), provided that the operator
$\mathcal{Q}$ in Eq. (\ref{MSE}) is replaced by the operator $\mathcal{Q}%
_{n^{\prime},n^{\prime\prime}}$.

\textbf{Three spin 1/2 system} - Partial disentanglement is explored below for
the relatively simple case of a system composed of three spin 1/2 particles.
For this case $d_{1}=d_{2}=d_{3}=2$, $d_{n}^{2}-1=3$, and the set of
generalized Gell-Mann matrices for the $n$'th spin is the set of Pauli
matrices $\left\{  2\hbar^{-1}\mathbf{S}_{n}\cdot\mathbf{\hat{x}},2\hbar
^{-1}\mathbf{S}_{n}\cdot\mathbf{\hat{y}},2\hbar^{-1}\mathbf{S}_{n}%
\cdot\mathbf{\hat{z}}\right\}  \dot{=}\{\lambda_{1}^{\left(  n\right)
},\lambda_{2}^{\left(  n\right)  },\lambda_{3}^{\left(  n\right)  }\}$, where
$\mathbf{S}_{n}=\left(  S_{nx},S_{ny},S_{nz}\right)  $ is the angular momentum
vector operator of the $n$'th spin (the symbol $\dot{=}$\ stands for matrix
representation), and $n\in\left\{  1,2,3\right\}  $. Consider a pure
normalized state $\left\vert \psi\right\rangle $ given by%
\begin{align}
\left\vert \psi\right\rangle  &  =q_{000}\left\vert 000\right\rangle
+q_{001}\left\vert 001\right\rangle +q_{010}\left\vert 010\right\rangle
+q_{011}\left\vert 011\right\rangle \nonumber\\
&  +q_{100}\left\vert 100\right\rangle +q_{101}\left\vert 101\right\rangle
+q_{110}\left\vert 110\right\rangle +q_{111}\left\vert 111\right\rangle
\;.\nonumber\\
&  \label{psi 3 spins}%
\end{align}
The ket vector $\left\vert \sigma_{3}\sigma_{2}\sigma_{1}\right\rangle $,
where $\sigma_{n}\in\left\{  0,1\right\}  $ and $n\in\left\{  1,2,3\right\}
$, represents an eigenvector of the matrix $(1-\lambda_{3}^{\left(  n\right)
})/2=\operatorname{diag}\left(  0,1\right)  $ with the eigenvalue $\sigma_{n}%
$, for $n=1$, $n=2$, and $n=3$. The state $\left\vert \psi\right\rangle $ is
characterized by the single spin Bloch vectors $\mathbf{k}_{n}=(\langle
\lambda_{1}^{\left(  n\right)  }\rangle,\langle\lambda_{2}^{\left(  n\right)
}\rangle,\langle\lambda_{3}^{\left(  n\right)  }\rangle)$, for $n=1$, $n=2$,
and $n=3$. The length of the $n$'th Bloch vectors $\mathbf{k}_{n}$, which is
denoted by $k_{n}=\left\vert \mathbf{k}_{n}\right\vert $, is bounded between
zero (fully entangled) and unity (fully disentangled, or fully separable).

For this three spin 1/2 system, the partial entanglements $\tau_{23}$ ,
$\tau_{31}$\ and $\tau_{12}$ are bounded between zero and unity, provided that
$\eta_{23}=\eta_{31}=\eta_{12}=1/3$ \cite{Schlienz_4396}. Disentanglement
between subsystems 1 and 2 is studied below using the modified Schr\"{o}dinger
equation (\ref{MSE}), with $\mathcal{H}^{{}}=0$, and with the operator
$\mathcal{Q}$ taken to be given by $\mathcal{Q}=\mathcal{Q}_{12}$ [see Eq.
(\ref{Q12}), and Eq. (8.163) of Ref. \cite{Buks_QMLN}]. For this case, the
time evolution governed by Eq. (\ref{MSE}) gives rise to monotonic decrease of
$\tau_{12}$ in time $t$.

\begin{figure}[ptb]
\begin{center}
\includegraphics[width=3.2in,keepaspectratio]{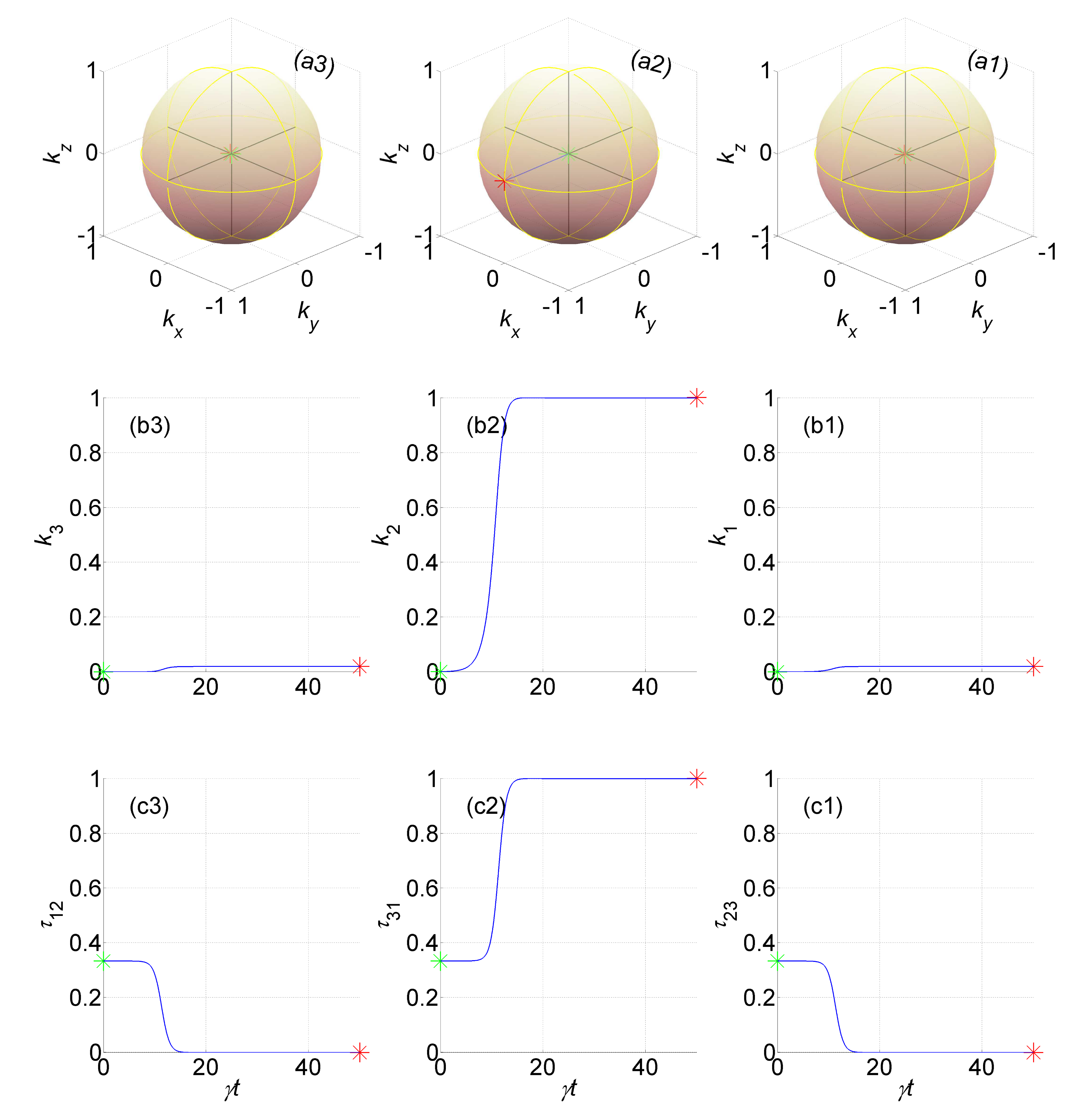}
\end{center}
\caption{{}Time evolution for $\left\vert \psi_{\mathrm{i} }\right\rangle
\simeq\left\vert \psi_{\mathrm{GHZ}}\right\rangle $. The initial state, which
is given by $\left\vert \psi_{\mathrm{i}}\right\rangle \sim\left\vert
\psi_{\mathrm{GHZ}}\right\rangle -10^{-5}i\left\vert \psi_{\mathrm{B}
,1}\left(  \pi\right)  \right\rangle -5.2\times10^{-4} i\left\vert
\psi_{\mathrm{B},2}\left(  \pi\right)  \right\rangle $, is represented by
green star symbols, and the final state at time $t=50/\gamma$ by red star
symbols.}%
\label{FigGHZT}%
\end{figure}

\begin{figure}[ptb]
\begin{center}
\includegraphics[width=3.2in,keepaspectratio]{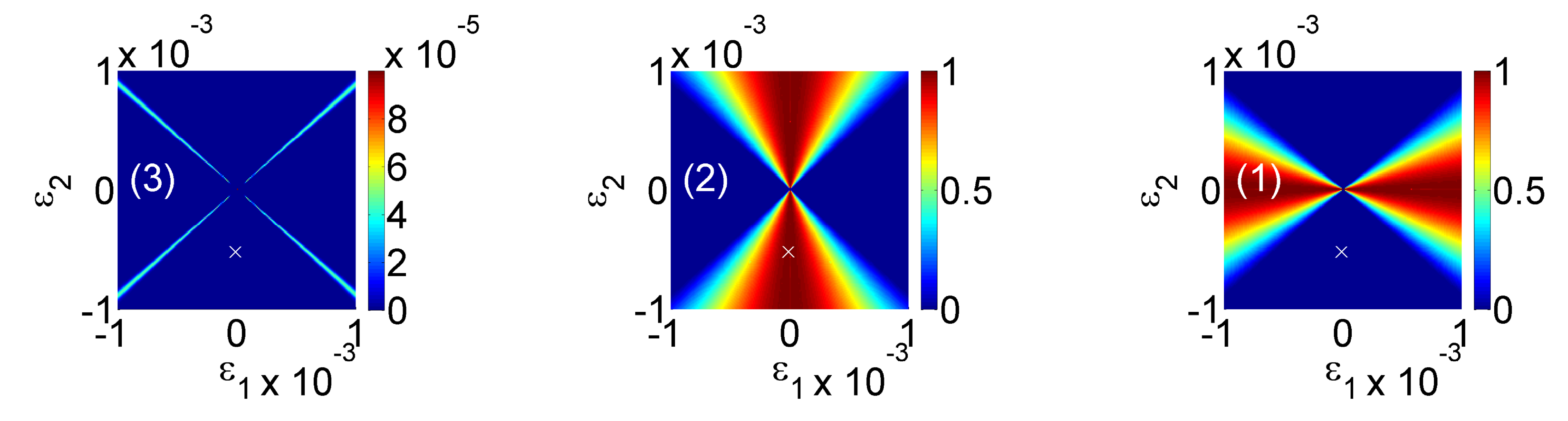}
\end{center}
\caption{{}Partial entanglements for $\left\vert \psi_{\mathrm{i}
}\right\rangle \sim\left\vert \psi_{\mathrm{GHZ}}\right\rangle +i\epsilon
_{1}\left\vert \psi_{\mathrm{B},1}\left(  \pi\right)  \right\rangle
+i\epsilon_{2}\left\vert \psi_{\mathrm{B},2}\left(  \pi\right)  \right\rangle
$. The values of $\tau_{23}$, $\tau_{31}$, and $\tau_{12}$ at the final time
$t=50/\gamma$ are shown in (1), (2) and (3), respectively, as a function of
$\epsilon_{1}$ and $\epsilon_{2}$. The overlaid white $\times$ symbol
represents the initial state $\left\vert \psi_{\mathrm{i}}\right\rangle $ used
for generating the plots in Fig. \ref{FigGHZT}.}%
\label{FigGHZM}%
\end{figure}

For the plots shown in Fig. \ref{FigGHZT} and Fig. \ref{FigGHZM} , the initial
state vector, which is denoted by $\left\vert \psi_{\mathrm{i}}\right\rangle
$, is close to the Greenberger Horne Zeilinger fully entangled state
$\left\vert \psi_{\mathrm{GHZ}}\right\rangle $, which is given by
\cite{Greenberger_69}%
\begin{equation}
\left\vert \psi_{\mathrm{GHZ}}\right\rangle =\frac{\left\vert 000\right\rangle
-\left\vert 111\right\rangle }{\sqrt{2}}\;.
\end{equation}
The plots in Fig. \ref{FigGHZT} exhibit time evolution from initial state
$\left\vert \psi_{\mathrm{i}}\right\rangle $ at time $t=0$ (labelled by green
star symbols), to final state $\left\vert \psi_{\mathrm{f}}\right\rangle $ at
time $t=50/\gamma$ (labelled by red star symbols). For the $n$'th spin, the
Bloch vector is shown in Fig. \ref{FigGHZT}(a$n$), the Bloch vector length
$k_{n}$ as a function of time in Fig. \ref{FigGHZT}(b$n$), and the
corresponding complement subsystem partial entanglement, i.e. $\tau_{23}$ for
$n=1$, $\tau_{31}$ for $n=2$, $\tau_{12}$ for $n=3$, as a function of time in
Fig. \ref{FigGHZT}(c$n$).

For the example shown in Fig. \ref{FigGHZT}, the initial state $\left\vert
\psi_{\mathrm{i}}\right\rangle $ is given by $\left\vert \psi_{\mathrm{i}%
}\right\rangle \sim\left\vert \psi_{\mathrm{GHZ}}\right\rangle +i\epsilon
_{1}\left\vert \psi_{\mathrm{B},1}\left(  \pi\right)  \right\rangle
+i\epsilon_{2}\left\vert \psi_{\mathrm{B},2}\left(  \pi\right)  \right\rangle
$, where $\epsilon_{1}=-10^{-5}$ and $\epsilon_{2}=-5.2\times10^{-4}$. The
symbol $\sim$ stands for equality up to normalization, i.e. $\left\vert
\psi_{\mathrm{i}}\right\rangle \sim\left\vert \psi^{\prime}\right\rangle $
implies that $\left\vert \psi_{\mathrm{i}}\right\rangle =\left\vert
\psi^{\prime}\right\rangle /\left\Vert \left\vert \psi^{\prime}\right\rangle
\right\Vert $. The states $\left\vert \psi_{\mathrm{B},n}\left(
\theta\right)  \right\rangle $, where $n\in\left\{  1,2,3\right\}  $, and
where $\theta$ is real, are two-spin Bell states given by%
\begin{align}
\left\vert \psi_{\mathrm{B},1}\left(  \theta\right)  \right\rangle  &
=\frac{\left\vert 000\right\rangle +e^{i\theta}\left\vert 110\right\rangle
}{\sqrt{2}}\;,\\
\left\vert \psi_{\mathrm{B},2}\left(  \theta\right)  \right\rangle  &
=\frac{\left\vert 000\right\rangle +e^{i\theta}\left\vert 101\right\rangle
}{\sqrt{2}}\;,\\
\left\vert \psi_{\mathrm{B},3}\left(  \theta\right)  \right\rangle  &
=\frac{\left\vert 000\right\rangle +e^{i\theta}\left\vert 011\right\rangle
}{\sqrt{2}}\;.
\end{align}
The overlaid white $\times$ symbol in Fig. \ref{FigGHZM} represents the
initial state $\left\vert \psi_{\mathrm{i}}\right\rangle $ used for generating
the plots in Fig. \ref{FigGHZT}. For this initial state $\left\vert
\psi_{\mathrm{i}}\right\rangle $, the final state $\left\vert \psi
_{\mathrm{f}}\right\rangle $ is approximately $\left\vert \psi_{\mathrm{f}%
}\right\rangle \simeq\left(  1/2\right)  \left(  \left\vert 000\right\rangle
+i\left\vert 010\right\rangle +i\left\vert 101\right\rangle -\left\vert
111\right\rangle \right)  $. The values of the partial entanglements
$\tau_{23}$, $\tau_{31}$, and $\tau_{12}$ at the final time $t=50/\gamma$ are
shown in Fig. \ref{FigGHZM} (1), (2) and (3), respectively, as a function of
$\epsilon_{1}$ and $\epsilon_{2}$.

\begin{figure}[ptb]
\begin{center}
\includegraphics[width=3.2in,keepaspectratio]{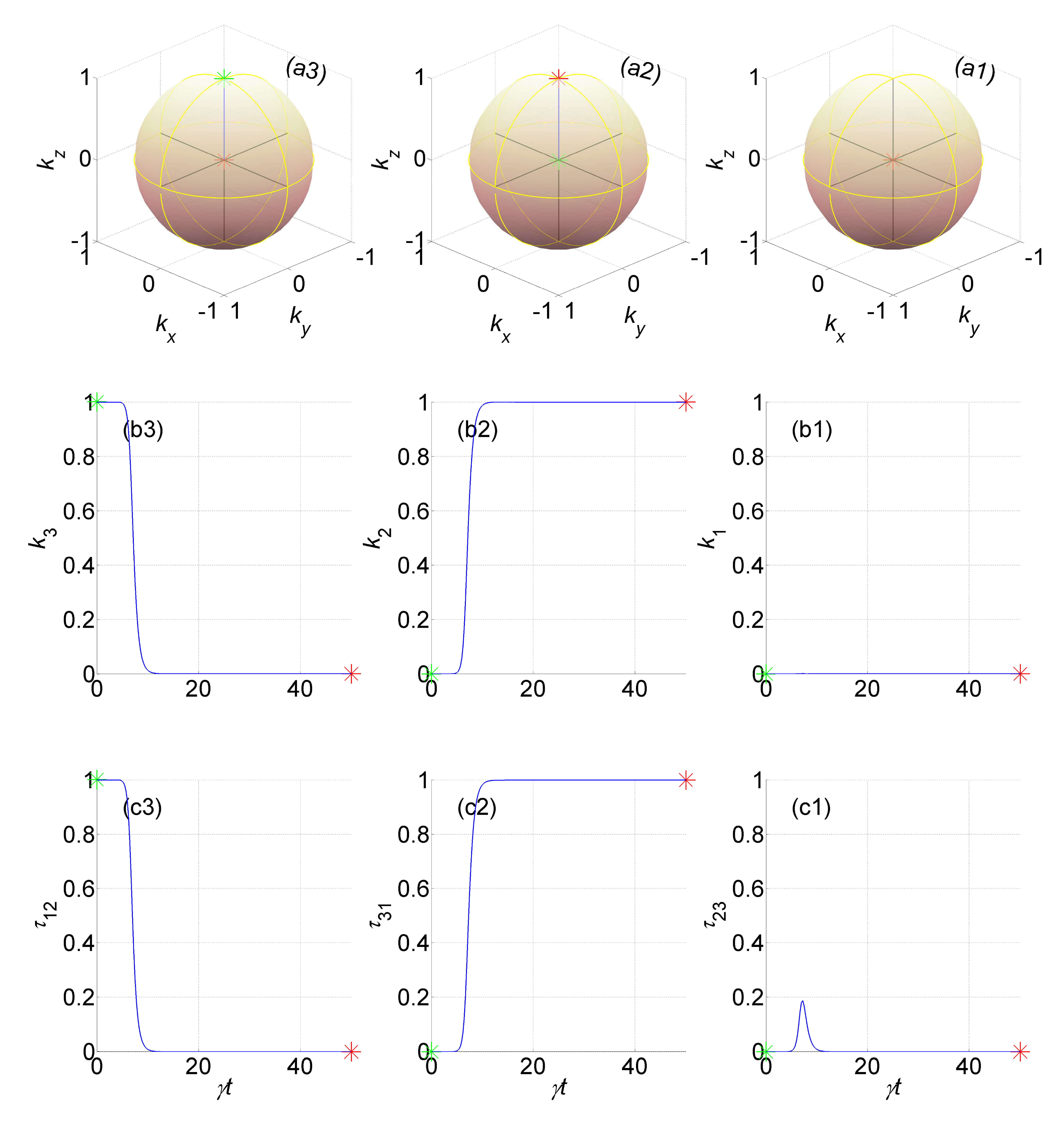}
\end{center}
\caption{Time evolution for $\left\vert \psi_{\mathrm{i}}\right\rangle
\simeq\left\vert \psi_{\mathrm{B},3}\left(  \pi\right)  \right\rangle $. The
initial state, which is given by $\left\vert \psi_{\mathrm{i}}\right\rangle
\sim\left\vert \psi_{\mathrm{B},3}\left(  \pi\right)  \right\rangle
+9\times10^{-5}i\left\vert \psi_{\mathrm{B},2}\left(  \pi\right)
\right\rangle $, is represented by green star symbols, and the final state at
time $t=50/\gamma$ by red star symbols.}%
\label{FigBS3T}%
\end{figure}

\begin{figure}[ptb]
\begin{center}
\includegraphics[width=3.2in,keepaspectratio]{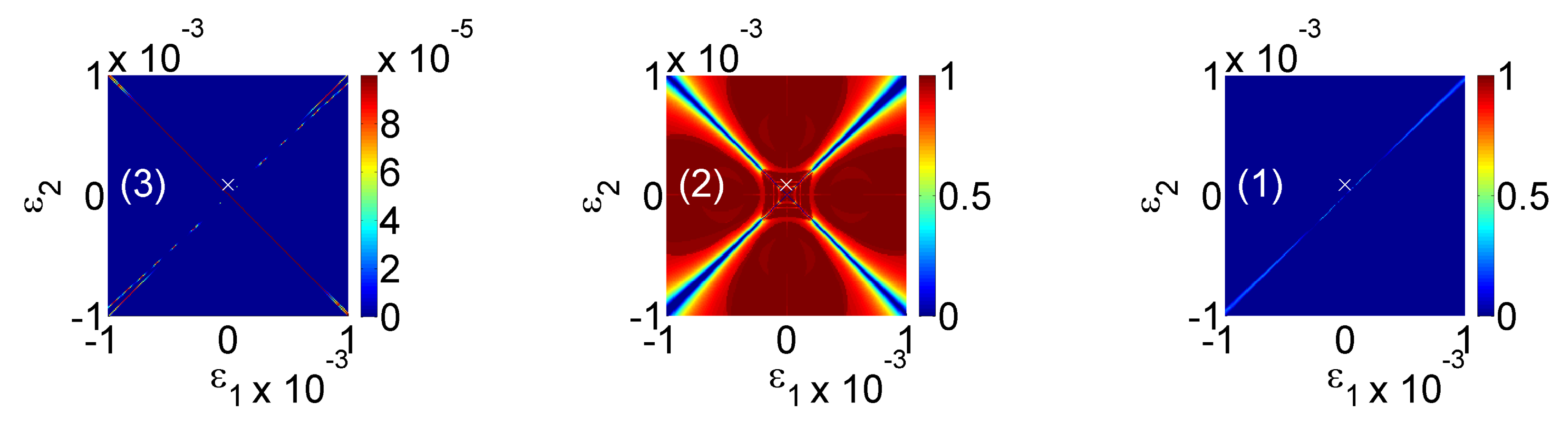}
\end{center}
\caption{Partial entanglements for $\left\vert \psi_{\mathrm{i}}\right\rangle
\sim\left\vert \psi_{\mathrm{B},3}\left(  \pi\right)  \right\rangle
+i\epsilon_{1}\left\vert \psi_{\mathrm{B},1}\left(  \pi\right)  \right\rangle
+i\epsilon_{2}\left\vert \psi_{\mathrm{B},2}\left(  \pi\right)  \right\rangle
$. The values of $\tau_{23}$, $\tau_{31}$, and $\tau_{12}$ at the final time
$t=50/\gamma$ are shown in (1), (2) and (3), respectively, as a function of
$\epsilon_{1}$ and $\epsilon_{2}$. The overlaid white $\times$ symbol
represents the initial state $\left\vert \psi_{\mathrm{i}}\right\rangle $ used
for generating the plots in Fig. \ref{FigBS3T}.}%
\label{FigBS3M}%
\end{figure}

As is demonstrated by Fig. \ref{FigGHZT}(c3) and Fig. \ref{FigBS3T}(c3),
generically, $\tau_{12}\rightarrow0$ in the limit $t\rightarrow\infty$, due to
the partial disentanglement generated by the term proportional to
$\mathcal{Q}_{12}-\left\langle \mathcal{Q}_{12}\right\rangle $ in Eq.
(\ref{MSE}). The set $V_{n}\subset H$ is a subset of the Hilbert space
$H=H_{3}\otimes H_{2}\otimes H_{1}$ containing all state vectors $\left\vert
\psi\right\rangle \in H$, for which $k_{n}=1$, where $n\in\left\{
1,2,3\right\}  $, i.e. the $n$'th spin is fully separable for all $\left\vert
\psi\right\rangle \in V_{n}$. Note that $V_{1}\cap V_{2}=V_{2}\cap V_{3}%
=V_{3}\cap V_{1}$ is the set of fully disentangled states (i.e. $k_{1}%
=k_{2}=k_{3}=1$), which is denoted by $V$. Let $\left\vert \psi_{\infty
}\right\rangle $ be the solution of Eq. (\ref{MSE}) in the limit
$t\rightarrow\infty$, for which $\tau_{12}=0$ (i.e. entanglement between spin
1 and spin 2 vanishes). Generically, either $\left\vert \psi_{\infty
}\right\rangle \subset V_{1}$ (for this case $k_{1}=1$, $k_{2}=k_{3}\leq1$,
$\tau_{12}=\tau_{31}=0$, and $\tau_{23}\geq0$ ), or $\left\vert \psi_{\infty
}\right\rangle \subset V_{2}$ (for this case $k_{2}=1$, $k_{3}=k_{1}\leq1$,
$\tau_{12}=\tau_{23}=0$, and $\tau_{31}\geq0$). Note that $\left\vert
\psi_{\infty}\right\rangle \subset V_{2}$ for both examples shown in Fig.
\ref{FigGHZT} and in Fig. \ref{FigBS3T}. For the general case, the entire
Hilbert space $H$ is divided into two basins of attraction, $B_{1}$ and
$B_{2}$. The basin $B_{n}$ is the set of all initial states $\left\vert
\psi_{\mathrm{i}}\right\rangle \in H$, for which in the limit $t\rightarrow
\infty$ the final state $\left\vert \psi_{\infty}\right\rangle \subset V_{n}$,
i.e. $k_{n}=1$, where $n\in\left\{  1,2\right\}  $.

As can be seen from Fig. \ref{FigGHZM}, the GHZ state vector $\left\vert
\psi_{\mathrm{GHZ}}\right\rangle $ lies on the separatrix between the basins
of attraction $B_{1}$ and $B_{2}$. The strong dependency of $\left\vert
\psi_{\infty}\right\rangle $ on the initial state $\left\vert \psi
_{\mathrm{i}}\right\rangle $, which becomes extreme in the vicinity of
$\left\vert \psi_{\mathrm{GHZ}}\right\rangle $, resembles the butterfly
effect. As is demonstrated below in Fig. \ref{FigBS3M}, a similar butterfly
effect occurs near the state vector $\left\vert \psi_{\mathrm{B},3}\left(
\pi\right)  \right\rangle $.

For the plots shown in Fig. \ref{FigBS3T} and Fig. \ref{FigBS3M} , the initial
state vector $\left\vert \psi_{\mathrm{i}}\right\rangle $ is given by
$\left\vert \psi_{\mathrm{i}}\right\rangle \sim\left\vert \psi_{\mathrm{B}%
,3}\left(  \pi\right)  \right\rangle +i\epsilon_{1}\left\vert \psi
_{\mathrm{B},1}\left(  \pi\right)  \right\rangle +i\epsilon_{2}\left\vert
\psi_{\mathrm{B},2}\left(  \pi\right)  \right\rangle $. For the example shown
in Fig. \ref{FigBS3T}, $\epsilon_{1}=0$ and $\epsilon_{2}=9\times10^{-5}$.
This initial state is indicated by the white $\times$ symbol in Fig.
\ref{FigBS3M}. The corresponding final state $\left\vert \psi_{\mathrm{f}%
}\right\rangle $ is approximately $\left\vert \psi_{\mathrm{f}}\right\rangle
\simeq\left\vert \psi_{\mathrm{B},2}\left(  -\pi/2\right)  \right\rangle $. As
can be seen from Fig. \ref{FigBS3M}, the butterfly effect occurs near the
state $\left\vert \psi_{\mathrm{B},3}\left(  \pi\right)  \right\rangle $,
which also (as the state $\left\vert \psi_{\mathrm{GHZ}}\right\rangle $) lies
on the separatrix between the basins of attraction $B_{1}$ and $B_{2}$.

\textbf{Summary} - The proposed modified Schr\"{o}dinger equation (\ref{MSE})
demonstrates a way to extend quantum mechanics to enable processes of
deterministic partial disentanglement in a multipartite system. Conflicts with
both principles of causality and separability can be avoided by postulating
that partial disentanglement between two given subsystems is active only
during the time when they interact \cite{Buks_2303_00697}. This assumption of
\textit{{local disentanglement}} implies that the disentanglement rate
$\gamma$ in Eq. (\ref{MSE}) vanishes during the time when subsystems are
decoupled. Further theoretical study is needed to determine whether quantum
mechanics can be self-consistently reformulated based on deterministic
dynamics. Alternative theoretical models for the process of quantum
measurement can be experimentally tested \cite{Buks_014421}.

\textbf{Acknowledgments} - A useful discussion with Eliahu Cohen is
acknowledged. This work was supported by the Israeli science foundation, and
the Israeli ministry of science.

\appendix

\section{The function $\tau\left(  \left\vert \psi\right\rangle \right)  $}

\label{App_tau}

For a given Hermitian operator $\mathcal{Q}^{{}}=\mathcal{Q}^{\dag}$, the
function $\tau\left(  \left\vert \psi\right\rangle \right)  $, which maps a
ket vector $\left\vert \psi\right\rangle $ to a real number $\tau$, is defined
by%
\begin{equation}
\tau\left(  \left\vert \psi\right\rangle \right)  \equiv\frac{\left\langle
\psi\right\vert \mathcal{Q}\left\vert \psi\right\rangle }{\left\langle
\psi\right.  \left\vert \psi\right\rangle }\;.
\end{equation}
Let $\Delta_{\tau}=\tau\left(  \left\vert \psi\right\rangle +\epsilon
\left\vert \delta\right\rangle \right)  -\tau\left(  \left\vert \psi
\right\rangle \right)  $, where $\epsilon$ is real, and $\left\vert
\delta\right\rangle $ is a ket vector. For $\left\langle \psi\right.
\left\vert \psi\right\rangle =1$ the following holds $\Delta_{\tau}%
=2\epsilon\operatorname{Re}\left\langle \delta\right.  \left\vert q_{\psi
}\right\rangle +O\left(  \epsilon^{2}\right)  $, where $\left\vert q_{\psi
}\right\rangle =\left(  \mathcal{Q}-\left\langle \psi\right\vert
\mathcal{Q}\left\vert \psi\right\rangle \right)  \left\vert \psi\right\rangle
$. Within the set of normalized ket vectors, the term $\operatorname{Re}%
\left\langle \delta\right.  \left\vert q_{\psi}\right\rangle $ is maximized
for $\left\vert \delta\right\rangle =\left\vert q_{\psi}\right\rangle
/\sqrt{\left\langle q_{\psi}\right.  \left\vert q_{\psi}\right\rangle }$.
Moreover, $\left\langle \psi\right.  \left\vert q_{\psi}\right\rangle =0$
provided that $\left\langle \psi\right.  \left\vert \psi\right\rangle =1$.
Based on these observations, the nonlinear term added to the Schr\"{o}dinger
equation (\ref{MSE}) is chosen to be proportional to $\left\vert q_{\psi
}\right\rangle $.

\bibliographystyle{ieeepes}
\bibliography{acompat,Eyal_Bib}

\end{document}